# Terahertz gas phase spectroscopy using a high finesse Fabry-Pérot cavity

Francis Hindle, Robin Bocquet, Arnaud Cuisset, Gaël Mouret,

Laboratoire de Physico-Chimie de l'Atmosphère, Université du Littoral-Côte d'Opale, 189 A Ave. Maurice Schumann 59140 Dunkerque, France.

Anastasiia Pienkina,

SATT-Nord, 25, Avenue Charles Saint-Venant - 59800 Lille, France

The achievable instrument sensitivity is a critical parameter for the continued development of THz applications. Techniques such as Cavity-Enhanced Techniques and Cavity Ring Down Spectroscopy have not yet been employed at THz frequency due to the difficulties to construct a high finesse Fabry-Pérot cavity. Here, we describe such a THz resonator based on a low-loss oversized corrugated waveguide with highly reflective photonic mirrors obtaining a finesse above 3000 around 620 GHz. These components enable a Fabry-Perot THz Absorption Spectrometer with an equivalent interaction length of one kilometer giving access to line intensities as low as $10^{-27}$ cm$^{-1}$/(molecule/cm$^2$) with a S/N ratio of 3. In addition, the intracavity optical power have allowed the Lamb-Dip effect to be studied with a low-power emitter, an absolute frequency accuracy better than 5 kHz can be easily obtained providing an additional solution for rotational spectroscopy.

## 1. INTRODUCTION

THz waves are able to probe a wide variety of molecular transitions, rotational energy levels of small polar compounds, or low energy vibrational levels of flexible molecules with active intra or intermolecular modes. In the gas phase and at low pressure, high-resolution THz spectroscopy has demonstrated its selectivity, thanks to Doppler broadenings of the rotational transitions which never exceeds tens of MHz, and the specificity of the low-frequency rovibrational signatures allowing to discriminate molecular compounds with nearby structures such as conformers or isomers [1,2]. Radio astronomers have used THz waves to complete a wide range of studies including unstable species such as radicals, cations and anions. The spectral windows defined by Herschel and SOFIA have provided additional impetus and lead to the detection of larger molecules fashionably referred to as 'prebiotic' [3]. THz spectroscopy with its strong rotational signatures is of interest to probe not only the earth's atmosphere but that of other planets containing more exotic molecular compositions and interstellar medium [4]. Applications such as, breath analysis [5] and environmental surveillance [6,7], amongst others [8,9], should be feasible if sufficient instrument sensitivity and spectral resolution can be obtained. Despite numerous advances observed at THz frequencies including THz Quantum Cascade Lasers [10–12], solid-state electronic devices [13], photonic conversions [14], detectors [15,16], this spectral region remains hindered by the lack of advanced technological system components available compared to the neighbouring microwave and infrared domains. The Fourier Transform instruments, as developed by Balle-Flygare, have proved particularly useful for pure rotational spectroscopy at centimeter and millimeter wavelengths especially when coupled with a pulsed molecular beam [17–19]. To reduce the acquisition time, chirped pulsed spectrometers have been developed recently and are now widely employed at microwave frequencies [20]. They are also presently being developed and exploited at higher frequencies with demonstrated instruments operating at frequencies up to 850 GHz [21–23]. Such instruments offer wideband measurement with a dynamic range (up to 50 dB) that can be readily matched to the intensity of the targeted lines. They however require two mm-wave multipliers and expensive high bandwidth electronics (arbitrary wave generator and data acquisition). In addition, at higher frequency the molecular dephasing time is shorter making these measurements more difficult. Nevertheless, chirped pulsed instruments in the mm-wave band are attracting interest for rotational spectroscopy [24] and chemical kinetics [25]. Above 300 GHz standard absorption spectroscopy is generally the only solution for high spectral resolution rotational spectroscopy. The achievable sensitivity is critical, a standard single pass measurement cell is typically limited to interaction lengths of several meters. Multiple pass cells provide higher sensitivities but are limited by a significant attenuation and generally require large volumes to reach distances typically not exceeding one hundred meters [26,27]. An alternative approach would be to adapt the techniques developed in the InfraRed domain, such as Cavity-Enhanced Absorption Spectroscopy (CEAS) and Cavity Ring Down Spectroscopy (CRDS) to the THz and submillimetre domain. Up to now the construction of a high finesse Fabry Pérot (FP) cavity has been the principal obstacle preventing this approach from succeeding. Here, we show a high finesse THz cavity and how it can be exploited to create a high spectral resolution Fabry-Perot THz Absorption Spectrometer (FP-TAS) with a kilometer equivalent interaction length working at room temperature without temperature stabilization.

## 2. THz CAVITY

Cavity based techniques used at microwave frequencies [28,29] (<300 GHz) and in the Infrared [30,31] (>10 THz) have never been successfully developed in the submillimetre and THz domains, the result being that this frequency region continues to be described as the so called "frequency gap". Some attempts confined to the lower frequencies, around 250 and 300 GHz yielded a limited finesse below 400 [32,33]. At higher frequency, around 600 GHz, a finesse of 220 has been demonstrated using a whispering-gallery mode in a spherical silicon resonator [34,35]. A FP cavity using two photonic mirrors has already been tested at 300 GHz but only achieving a limited finesse [36]. Unlike in the infrared band, the absence of highly reflective spherical dielectric mirrors to confine the THz radiation in a stable cavity is a real obstacle. All losses, such as diffraction that can be significant at long wavelength, attenuation, and limited reflection mirrors have hampered the development of a high finesse FP THz cavities. The global scheme of the FP-TAS, is shown on Fig. 1; our high finesse FP cavity is made with a low-loss oversized corrugated waveguide designed to work around 600 GHz, along with high reflectivity 1D photonic mirrors. The circular waveguide is electroformed copper with gold plating supplied by Thomas Keating Company. It is intended for high power millimeter-wave applications such plasma heating experiments that cannot tolerate anything more than a low-level of losses. Such a structure is achieved when its radius is larger than wavelength, with a depth corresponding to quarter of the targeted wavelength [37–39]. The 1D-photonic mirrors are the second key element of this FP-TAS, constructed from a series of four high resistive silicon wafers (>8k$\Omega$.cm) selected for their negligible absorption. The wafers were separated by three spacers, the thicknesses of the wafers and spacers were selected to have a maximum reflectivity band around 620 GHz (see section 3).

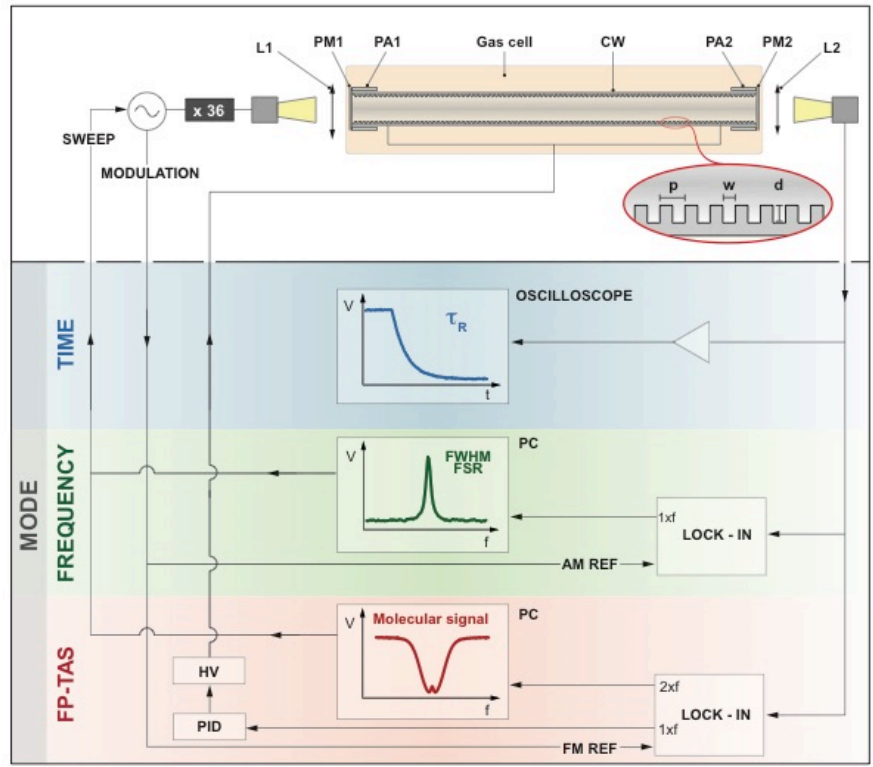

**Fig. 1. Fabry-Perot THz cavity system overview with three operational modes.** The emitter is an amplified multiplier chain (x36) covering 440-660 GHz driven by a microwave synthesizer referenced to a GPS time signal. The synthesizer is referenced to a GPS timing signal providing a frequency accuracy of $10^{-11}$ when measured over 1 second. The phase noise at the output of the frequency multiplier operating at 620 GHz is -63 dBc/Hz at 1 kHz from the carrier frequency. A TPX (Polymethylpentene) lens L1 (f=25 mm) is used to couple the free space THz emission to the corrugated waveguide CW. Two 1-D photonic mirrors PM1 and PM2 close the cavity, one at each end of the corrugated waveguide. Each photonic mirror is mounted on a piezo actuator PA1 and PA2, enabling fine tuning of the cavity length over at least 250 micrometers to ensure complete coverage. A second TPX lens L2 (f=25 mm) collects the THz emission at the cavity output and focuses it on a Zero Bias Schottky Detector Diode (ZBD WR1.5). The corrugated waveguide is 48 cm long, with internal diameters of 20.54 mm. The internal corrugations have a pitch of p = 166 $\mu$m, while the groves are w = 83 $\mu$m wide and d=125 $\mu$m deep. **Time mode**, the cavity output signal is amplified and recorded by an oscilloscope while the source is extinguished giving direct access to the cavity ring-down time $\tau_R$. **Frequency mode,** the THz source frequency is scanned and the cavity response measured using a lock-in detection and amplitude modulation of the source. The cavity mode linewidths (FWHM) and Free Spectral Range (FSR) are directly obtained in the frequency domain. **Fabry-Perot THz Absorption Spectrometer (FP-TAS) mode,** the THz source is frequency modulated and the cavity output measured by lock-in detection. The first harmonic (1xf) is used as an error signal, a cavity mode is locked to the frequency of the THz source using a Proportional, Integrator, Derivative (PID) control loop that feeds a High Voltage (HV) power supply. The second harmonic (2xf) provides a sensitive molecular signal as the source frequency and cavity scan together. The entire cavity assembly is placed in a pressure-controlled gas cell equipped with Teflon windows.

A free space $TEM_{00}$ Gaussian beam with around 50 $\mu$W of power is coupled to the lowest order hybrid electric mode ($HE_{11}$) of the circular oversized corrugated waveguide and collected at the output of the waveguide by means of two TPX lenses. This fundamental mode presents the lowest losses and must be used in such an application. Ideally, the free space $TEM_{00}$ mode can be coupled to the $HE_{11}$ mode with an efficiency of 98 %, and conversely this latter is projected almost entirely onto a free space Gaussian mode at the exit of the waveguide [40]. The linearly polarized $HE_{11}$ is also well suited to the Zero Bias Schottky diode detector located in a rectangular waveguide and coupled to a horn antenna. We are also very mindful not to introduce any sources of mode conversion which may be caused by misalignments, any supplemental losses rapidly deteriorate the properties of the FP cavity.

Several related figures of merit such as finesse $F$, quality factor $Q$ or the ring-down time $\tau_R$ may be employed to characterize the cavity performance. They all give access to the total cavity losses including the mirror reflectivity and waveguide losses. The traceable high frequency electronic source referenced onto a GPS signal, provides a stable frequency that can be easily controlled which facilitates the cavity characterization. Although it is not generally the case for laser-based experiments, when using an electronic source, the finesse and the photon life time can be measured with the same setup without any additional complexity [41,42]. Once enough THz power has been injected into the FP cavity at a resonance frequency, the emitter is rapidly switched off, the exponential decay is recorded and the ring down time $\tau_R$ obtained. As shown in Fig. 2a, it has been evaluated at 1.571 +/- 0.004 µsec for a resonance centered at 636.6 GHz without any gas in the cavity (residual pressure is less than 1 µbar). Alternatively, by measuring the transmitted power of the FP cavity as function of frequency, a succession of maxima is observed (Fig. 2b). The Free Spectral Range (*FSR*), and FWHM line-widths *(Δν)* give access to the cavity finesse $F$, defined by *F=FSR/(Δν)* where *FSR* is the frequency difference between two subsequences resonances, measured around 311.7 MHz in the present demonstration. Values of $F$ between 3000 and 3500 are clearly demonstrated, corresponding to a quality factor between 6 and $7 \cdot 10^6$. This cavity by far outperforms all previous attempts in this frequency band, and so unlocks new possibilities in the "THz frequency gap". The relation between Δν, used to calculate the finesse $F$, and ring-down time $\tau_R$ defined by $\tau_R=1/2\pi\Delta\nu$ is readily verified, enabling straightforward measurement in either the frequency or temporal domains.

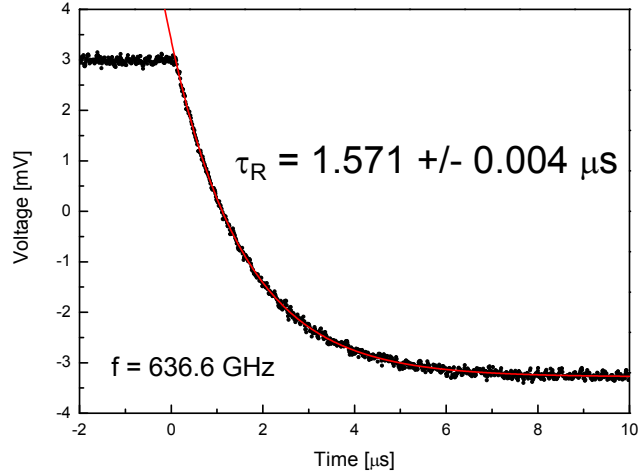

**Fig. 2a. Time mode**. Ring-down signal of the cavity recorded after the extinction of the THz source. Black points measured data. Red line fitted exponential curve. THz source operating at 636.6 GHz, cavity length is tuned to match the cavity mode to the source frequency. Measurements are performed with an evacuated gas cell, residual pressure is less than 1 µbar. Measured trace is an average of 60 000 acquisitions obtained in less than 5 minutes.

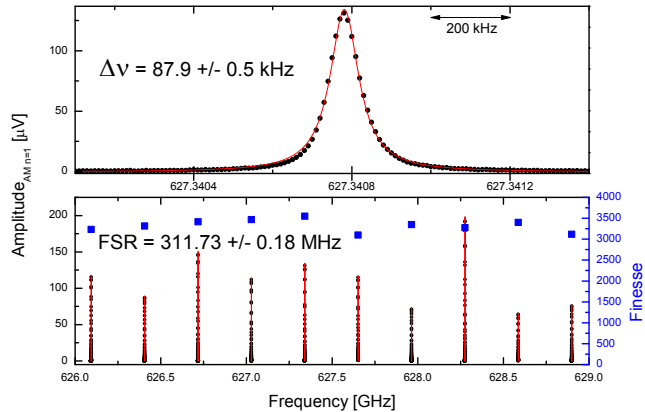

**Fig. 2b. Frequency mode**. Successive cavity modes observed for a fixed cavity length. Black circles are measured data. Red lines are data fitted using a Lorentzian function. Blue squares show the Finesse *F=FSR/Δν*. Upper pane shows a detailed plot of the resonance around 627.3408 GHz. These data were measured with an amplitude modulation of 10 kHz, a frequency step of 10 kHz and an integration time of 30ms/point. The uncertainty of the FSR values is dominated by the temperature fluctuation, a variation of $10^{-3}$ K shifts frequency position of a given resonance by 10 kHz.

The cavity finesse $F$ expressed in terms of the mirror reflectivity, and waveguide losses is given by:

$$F = \frac{\pi\sqrt{R \cdot e^{-\alpha \cdot L}}}{1 - R \cdot e^{-\alpha \cdot L}} \quad (1)$$

$R$ is the reflectivity of 1D photonics mirrors, $\alpha$ the waveguide losses by unit length, and $L$ the length of the FP cavity. Experimentally the average finesse achieved is 3200 +/- 20, corresponding to total losses of $R \cdot e^{-\alpha \cdot L} = 0.99902$. It remains difficult to accurately separate the contributions from $R$ and $\alpha$, both of which may limit maximum achievable finesse. Comparison of the cavities using 2 different waveguide lengths would be the best approach. However, the manufacturing difficulties to produce identical waveguides 2 or 3 times longer could not be overcome. Nevertheless, by calculating the reflectivity of mirrors (see section 3) by including absorption of Silicon we may evaluate $R$ and $\alpha$. Even though the high resistivity silicon is one of the most transparent materials in the THz domain, its weak absorption cannot be ignored in the present application. Significant discrepancies are present in the available literature for this parameter. The losses of the Silicon wafer used to implement the 1-D photonics mirrors will be considered to be less than 0.02 cm$^{-1}$ as proposed in Ref. [43]. This value corresponds to a calculated maximum reflectivity of 99.95% yielding a finesse of 6000 with zero waveguide losses. Based on this hypothesis, the finesse is predominantly limited by the waveguide losses that are estimated to be around 0.004 dB/m or $10^{-5}$ cm$^{-1}$. The present estimated losses are in reasonable agreement with the calculated ($7 \cdot 10^{-6}$ cm$^{-1}$) and measured ($4 \cdot 10^{-5}$ cm$^{-1}$) values for similar waveguides working at lower frequency [44–46]. The losses of such a waveguide are known to scale
inversely with the cube of the waveguide radius [37]. A comparison can be made with a cavity composed of a 18 cm long waveguide with an internal diameter of 12.7 mm and the two 1D photonics mirrors, yielding a finesse of 2500 around 620 GHz. Using the same silicon losses, we have estimated the waveguide losses to be 0.017dB/m or $4.2 \cdot 10^{-5}$ cm$^{-1}$ in agreement with the expected variation. The accuracy of these estimations is directly dependent on the uncertainty of the silicon losses. If the silicon is more transparent than 0.02 cm$^{-1}$ the waveguide losses have been underestimated, or conversely if the silicon is less transparent the waveguide losses are overestimated.

## 3. THz MIRRORS

Low loss, ultrahigh reflectance homemade mirrors have been designed and fabricated by a periodically layering of quarter-wave optical thicknesses of silicon and vacuum in the usual Bragg configuration. These multilayers structures have already been investigated by other groups but never used in a context of high-resolution THz spectroscopy and also associated to common difficulties to measure accurately the reflectivity [47–50]. Mirrors composed of three layers have been previously used to form a FP cavity, the published values of silicon losses allow the achieved reflectivity for these mirrors to be estimated at 99.7% [36]. To obtain a higher reflectivity we used four double side polished high resistivity silicon wafers (> 8kΩ.cm) separated by three vacuum spacers to form the supermirrors. The spacers are held under the same vacuum as the sample environment. Silicon is one of lowest loss materials in THz with a constant refraction index of 3.4175 over a relatively large bandwidth that optimized the index contrast and thus reflectivity. Spacers are made by silicon wafers with a hole of 21 mm diameter in it, superior to internal diameter of the corrugated waveguide. The supermirror design is based on a matrix method to target a maximum of reflectivity, estimated around to be 99.95 %, at 620 GHz including an absorption of 0.02 cm$^{-1}$ for silicon wafers. To prevent the need for custom manufacturing, available high resistive silicon wafers with thicknesses of 185 µm and 375 µm were selected for the spacers and discs respectively. The wafers have a thickness tolerance of ± 5 µm. Such 1D photonic mirrors require very high resistive silicon material and can be easily and cheaply designed for other target frequencies due to the sub-mm dimensions.

## 4. FABRY-PEROT THz ABSORPTION SPECTROMETER

An important difference with the IR region is the Doppler limited linewidth, which is much narrower at THz frequencies. A feedback control loop was used to correct the cavity resonant frequency to match the frequency of the source. This maintains the maximum transmission of the FP cavity and allows the mode to be tuned with the frequency of the source across the molecular line profile to be scanned. The correction signal is applied to a piezo-mechanic actuator changing the distance between the mirrors. The source is frequency modulated and the detected signal processed by a lock-in amplifier giving access to its different harmonics. The first harmonic is used as a signal discriminator to drive a Proportional Integrator Derivative (PID) regulator and a high voltage supply that controls the position of 1D photonics mirrors and thus the cavity length. The second harmonic of the detected signal indicates the presence of absorbing compounds. When a molecular absorption is introduced in the cavity, both the finesse and peak transmitted intensity are decreased. The second harmonic signal is very sensitive in this case as both effects will contribute to a reduction of the measured signal. Carbonyl sulfide (OCS) was selected to assess the performance and potential of the FP-TAS for high resolution spectroscopy. Our prototype proved to be too sensitive to measure absorption of the principal isotopologue, we focused our attention on the isotopologue $^{18}O^{13}C^{32}S$ with a natural abundance of 21 ppm. The molecular signal is clearly visible in Fig. 3 at 636.604 930 GHz corresponding to the J"=56 ←J'=55 rotational transition of $^{18}O^{13}C^{32}S$, with an intensity of $2.8 \cdot 10^{-26}$ cm$^{-1}$/(molecule/cm$^2$) at 300 K taking into account its natural abundance [51,52]. The signal to noise ratio of molecular signal here is estimated to be greater than 60. Such an absorption reduced also the ring down time of 0.1 µsec at the maximum of the absorption. The equivalent interaction length $L_{eq}=(2 \cdot F \cdot L)/\pi$, is estimated at around one kilometer in the present demonstration. Such an enhancement of sensitivity is required to clearly detect this target absorption line without using cryogenically cooled detector and with only ≈ 50 µW of power.

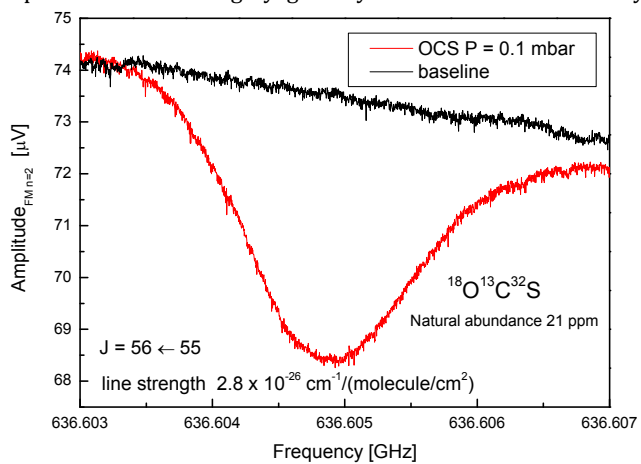

**Fig. 3. Fabry-Perot THz Absorption Spectrometer (FP-TAS) mode.** Red line is the second harmonic of the frequency modulated THz wave at the cavity output with a pressure of 100 µbar of OCS. The transition J = 56 ← 55 of the isotopologue $^{18}O^{13}C^{32}S$ is clearly recorded. Black line is the baseline measurement under identical conditions with an empty gas cell. Integration time 200ms/point, frequency step 2 kHz, frequency modulation depth 90 kHz.

One of the most successful applications of THz rotational spectroscopy remains the study of terrestrial, planetary or cometary atmospheres and the interstellar medium. New astronomical telescopes such as Hershel Space Observatory, Stratospheric Observatory for Infrared Astronomy (Sofia) and Atacama Large Millimetre Array (Alma) have been deployed. Such facilities allow unprecedented possibilities for submillimetre and THz wave observations, they require very accurate frequency lines transitions to ensure an efficient identification [53]. The exact knowledge of molecular parameters is also very important, especially for the validation and calibration of theoretical models. Sub-Doppler spectral resolution techniques such as saturated absorption spectroscopy or Lamb-dip effect, are valuable approaches to greatly improve the accuracy with which the centre frequencies of rotational absorption lines can be measured. Moreover, different hyperfine structures may be resolved below the doppler limit. Such an approach is routinely employed in the millimetre wave range and generally requires extremely powerfull sources, cryogenic detectors, particularly long measurements times, and strong molecular absorption lines. The high finesse FP cavity ensures a high intracavity power enabling the simple observation of the Lamb-dip effect around 626 GHz, even for relatively weak rotational absorption lines of minor isotopologues, as illustrated in Fig. 4. The centre of the Lamb-dip was determined by fitting 10 spectra recorded at different pressures between 5 and 20 µbar using a Gaussian profile. A mean frequency value of 626 780 174 974 Hz was obtained with a standard deviation of 890 Hz for the J''=55 ← J'=54 rotational transition of $^{18}O^{12}C^{32}S$. We note a discrepancy of 48 kHz with the tabulated value available in the Cologne Database for Molecular Spectroscopy [51,54]. Depending on the available signal to noise ratio, an absolute frequency accuracy between 1 and 5 kHz can be easily obtained by use of this FP-TAS. The width of the observed lamb dip effect is of the same order compared to the cavity mode linewidth.

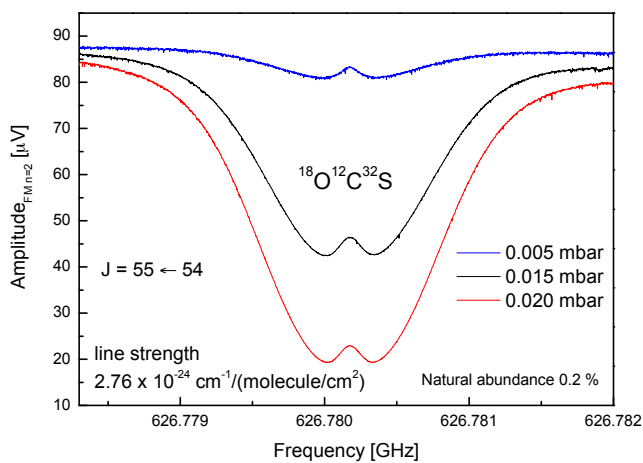

**Fig. 4. Fabry-Perot THz Absorption Spectrometer (FP-TAS) mode.** Lamb dip effect observed at different pressure on the J=55←54 rotational transition of $^{18}O^{12}C^{32}S$, recorded with an integration time of 100 ms/point, a frequency step of 2 kHz and a frequency modulation depth of 72 kHz

## 5. DISCUSSION & CONCLUSION

While the performance of THz components such as sources, detectors and mixers have improved over the last 20 years, high-resolution THz investigations are still hampered by the lack of sensitive cavity based techniques compared to other spectral domains. Our FP-TAS with an km interaction length provides a straightforward solution that can be easily replicated, it represents a critical step for the application of THz waves. The FP-TAS is clearly able to measure transitions with intensities of about $10^{-27}$cm$^{-1}$/(molecule/cm$^2$) with a S/N ratio of 3 around 620 GHz. This first prototype works at room temperature (without temperature stabilization), uses standard commercially available emitters and detectors, it can be easily adapted to various other spectral bands. Presently, the tunability of the spectrometer is limited by properties of the 1D-photonic mirrors, which cover a range from 600 to 650 GHz while maintaining a finesse greater than 10% of the maximum value. The coverage of a single mirror may be increased by using broadband components such as wire grid polarizers that are able to work over a wide range with a slight reduction of finesse. It also provides the opportunity to characterize highly reflective mirrors and/or very low losses waveguides which are otherwise particularly difficult to measure.

The available dynamic range is presently limited to about 30 dB, this may be extended by replacing the room temperature Zero Bias Schottky Diode Detector by a very low noise cryogenic detector, also improving the system sensitivity. This THz spectrometer combines high selectivity, and perfect frequency control, so simultaneously benefits from the advantages of Cavity Ring-Down Spectroscopy and Cavity Enhanced Absorption Spectroscopy, providing an additional solution for rotational spectroscopy above 300 GHz. The THz FP cavity now enables alternative approaches like Noise-Immune Cavity-Enhanced Optical-Heterodyne Molecular Spectroscopy (NICE-OHMS) to be pursued, or cavity-enhanced frequency comb Fourier transform spectroscopy and then would help to close the terahertz technology gap between microwave electronics and infrared photonics [55,56]. The FP-TAS should allow the quantitative analysis of light molecules contained in complex mixtures, for example $N_2O$, $NO_2$, $CH_3CN$ et $CH_3Cl$ all display maximum line strengths between 500 and 700 GHz at room temperature.

**Funding**. The authors would like to acknowledge the financial support of this work by the French Agence Nationale de la Recherche project Original Sub-millimeter Chirped pulse instrumentation for Astrochemical Reactivity (OSCAR) under contract number ANR-15-CE29-0017, and the European Fund for Regional Economic Development through the TERAFOOD project (INTERREG V FR-WA-VL 1.2.11).

**Acknowledgment**. We thank Dr. Marc Fourmentin (LPCA) for the preparation of the figures.

## REFERENCES

1. F. C. De Lucia, "The submillimeter: A spectroscopist's view," Journal of Molecular Spectroscopy **261**, 1–17 (2010).
2. Martin Quack, Frederic Merkt, *Handbook of High-Resolution Spectroscopy* (John Wiley & Sons, 2011).
3. Thaddeus P, "The prebiotic molecules observed in the interstellar gas," Philosophical Transactions of the Royal Society B: Biological Sciences **361**, 1681–1687 (2006).
4. Th. Encrenaz, B. Bézard, J. Crovisier, A. Coustenis, E. Lellouch, S. Gulkis, and S. K. Atreya, "Detectability of molecular species in planetary and satellite atmospheres from their rotational transitions," Planetary and Space Science **43**, 1485–1516 (1995).
5. L. W. Hrubesh and M. W. Droege, "Pure-rotational spectrometry: a vintage analytical method applied to modern breath analysis," J. Breath Res. **7**, 037105 (2013).

6. D. Bigourd, A. Cuisset, F. Hindle, S. Matton, E. Fertein, R. Bocquet, and G. Mouret, "Detection and quantification of multiple molecular species in mainstream cigarette smoke by continuous-wave terahertz spectroscopy," Opt. Lett., OL **31**, 2356–2358 (2006).
7. C. F. Neese, I. R. Medvedev, G. M. Plummer, A. J. Frank, C. D. Ball, and F. C. D. Lucia, "Compact Submillimeter/Terahertz Gas Sensor With Efficient Gas Collection, Preconcentration, and ppt Sensitivity," IEEE Sensors Journal **12**, 2565–2574 (2012).
8. F. Hindle, L. Kuuliala, M. Mouelhi, A. Cuisset, C. Bray, M. Vanwolleghem, F. Devlieghere, G. Mouret, and R. Bocquet, "Monitoring of food spoilage by high resolution THz analysis," Analyst **143**, 5536–5544 (2018).
9. A. Roucou, I. Kleiner, M. Goubet, S. Bteich, G. Mouret, R. Bocquet, F. Hindle, W. L. Meerts, and A. Cuisset, "Towards the Detection of Explosive Taggants: Microwave and Millimetre-Wave Gas-Phase Spectroscopies of 3-Nitrotoluene," ChemPhysChem **19**, 1056–1067 (2018).
10. R. Köhler, A. Tredicucci, F. Beltram, H. E. Beere, E. H. Linfield, A. G. Davies, D. A. Ritchie, R. C. Iotti, and F. Rossi, "Terahertz semiconductor-heterostructure laser," Nature **417**, 156 (2002).
11. S. Barbieri, P. Gellie, G. Santarelli, L. Ding, W. Maineult, C. Sirtori, R. Colombelli, H. Beere, and D. Ritchie, "Phase-locking of a 2.7-THz quantum cascade laser to a mode-locked erbium-doped fibre laser," Nature Photonics **4**, 636–640 (2010).
12. C. Sirtori, S. Barbieri, and R. Colombelli, "Wave engineering with THz quantum cascade lasers," Nature Photonics **7**, 691–701 (2013).
13. A. Maestrini, I. Mehdi, J. V. Siles, J. S. Ward, R. Lin, B. Thomas, C. Lee, J. Gill, G. Chattopadhyay, E. Schlecht, J. Pearson, and P. Siegel, "Design and Characterization of a Room Temperature All-Solid-State Electronic Source Tunable From 2.48 to 2.75 THz," IEEE Transactions on Terahertz Science and Technology **2**, 177–185 (2012).
14. A. D. J. F. Olvera, H. Lu, A. C. Gossard, and S. Preu, "Continuous-wave 1550 nm operated terahertz system using ErAs:In(Al)GaAs photo-conductors with 52 dB dynamic range at 1 THz," Opt. Express, OE **25**, 29492–29500 (2017).
15. A. Luukanen, E. N. Grossman, A. J. Miller, P. Helisto, J. S. Penttila, H. Sipola, and H. Seppa, "An Ultra-Low Noise Superconducting Antenna-Coupled Microbolometer With a Room-Temperature Read-Out," IEEE Microwave and Wireless Components Letters **16**, 464–466 (2006).
16. W. Zhang, P. Khosropanah, J. R. Gao, E. L. Kollberg, K. S. Yngvesson, T. Bansal, R. Barends, and T. M. Klapwijk, "Quantum noise in a terahertz hot electron bolometer mixer," Appl. Phys. Lett. **96**, 111113 (2010).
17. T. J. Balle, E. J. Campbell, M. R. Keenan, and W. H. Flygare, "A new method for observing the rotational spectra of weak molecular complexes: KrHCl," J. Chem. Phys. **72**, 922–932 (1980).
18. I. Merke and H. . Dreizler, "A Molecular Beam Fourier Transform Microwave Spectrometer in the Range 26.5 to 40 GHz. Tests of Performance and Analysis of the D-and 14N-Hyperfine Structure of Methylcyanide-d1," Zeitschrift für Naturforschung A **49**, (1994).
19. D. J. Nemchick, B. J. Drouin, M. J. Cich, T. Crawford, A. J. Tang, Y. Kim, T. J. Reck, E. T. Schlecht, M.-C. F. Chang, and G. Virbila, "A 90-102 GHz CMOS based pulsed Fourier transform spectrometer: New approaches for in situ chemical detection and millimeter-wave cavity-based molecular spectroscopy," Review of Scientific Instruments **89**, 073109 (2018).
20. G. G. Brown, B. C. Dian, K. O. Douglass, S. M. Geyer, S. T. Shipman, and B. H. Pate, "A broadband Fourier transform microwave spectrometer based on chirped pulse excitation," Review of Scientific Instruments **79**, 053103 (2008).
21. A. L. Steber, B. J. Harris, J. L. Neill, and B. H. Pate, "An arbitrary waveform generator based chirped pulse Fourier transform spectrometer operating from 260 to 295GHz," Journal of Molecular Spectroscopy **280**, 3–10 (2012).
22. J. L. Neill, B. J. Harris, A. L. Steber, K. O. Douglass, D. F. Plusquellic, and B. H. Pate, "Segmented chirped-pulse Fourier transform submillimeter spectroscopy for broadband gas analysis," Opt. Express, OE **21**, 19743–19749 (2013).
23. G. B. Park, A. H. Steeves, K. Kuyanov-Prozument, J. L. Neill, and R. W. Field, "Design and evaluation of a pulsed-jet chirped-pulse millimeter-wave spectrometer for the 70–102 GHz region," J. Chem. Phys. **135**, 024202 (2011).
24. N. Wehres, J. Maßen, K. Borisov, B. Schmidt, F. Lewen, U. U. Graf, C. E. Honingh, D. R. Higgins, and S. Schlemmer, "A laboratory heterodyne emission spectrometer at submillimeter wavelengths," Phys. Chem. Chem. Phys. **20**, 5530–5544 (2018).
25. C. Abeysekera, B. Joalland, N. Ariyasingha, L. N. Zack, I. R. Sims, R. W. Field, and A. G. Suits, "Product Branching in the Low Temperature Reaction of CN with Propyne by Chirped-Pulse Microwave Spectroscopy in a Uniform Supersonic Flow," J. Phys. Chem. Lett. **6**, 1599–1604 (2015).
26. J. S. Melinger, Y. Yang, M. Mandehgar, and D. Grischkowsky, "THz detection of small molecule vapors in the atmospheric transmission windows," Opt. Express, OE **20**, 6788–6807 (2012).
27. A. I. Meshkov and F. C. De Lucia, "Broadband absolute absorption measurements of atmospheric continua with millimeter wave cavity ringdown spectroscopy," Review of Scientific Instruments **76**, 083103 (2005).
28. D. A. Helms and W. Gordy, ""Forbidden" rotational spectra of symmetric-top molecules: PH3 and PD3," Journal of Molecular Spectroscopy **66**, 206–218 (1977).
29. M. A. Koshelev, I. I. Leonov, E. A. Serov, A. I. Chernova, A. A. Balashov, G. M. Bubnov, A. F. Andriyanov, A. P. Shkaev, V. V. Parshin, A. F. Krupnov, and M. Y. Tretyakov, "New Frontiers in Modern Resonator Spectroscopy," IEEE Transactions on Terahertz Science and Technology **8**, 773–783 (2018).
30. B. Bernhardt, A. Ozawa, P. Jacquet, M. Jacquey, Y. Kobayashi, T. Udem, R. Holzwarth, G. Guelachvili, T. W. Hänsch, and N. Picqué, "Cavity-enhanced dual-comb spectroscopy," Nature Photonics **4**, 55–57 (2010).
31. D. Romanini, A. A. Kachanov, N. Sadeghi, and F. Stoeckel, "CW cavity ring down spectroscopy," Chemical Physics Letters **264**, 316–322 (1997).
32. B. Alligood DePrince, B. E. Rocher, A. M. Carroll, and S. L. Widicus Weaver, "Extending high-finesse cavity techniques to the far-infrared," Review of Scientific Instruments **84**, 075107 (2013).
33. R. Braakman and G. A. Blake, "Principles and promise of Fabry–Perot resonators at terahertz frequencies," Journal of Applied Physics **109**, 063102 (2011).
34. D. W. Vogt and R. Leonhardt, "Fano resonances in a high-Q terahertz whispering-gallery mode resonator coupled to a multi-mode waveguide," Opt. Lett., OL **42**, 4359–4362 (2017).
35. D. W. Vogt and R. Leonhardt, "Ultra-high Q terahertz whispering-gallery modes in a silicon resonator," APL Photonics **3**, 051702 (2018).
36. T. Chen, P. Liu, J. Liu, and Z. Hong, "A terahertz photonic crystal cavity with high Q-factors," Appl. Phys. B **115**, 105–109 (2014).
37. E. J. Kowalski, D. S. Tax, M. A. Shapiro, J. R. Sirigiri, R. J. Temkin, T. S. Bigelow, and D. A. Rasmussen, "Linearly Polarized Modes of a Corrugated Metallic Waveguide," IEEE Transactions on Microwave Theory and Techniques **58**, 2772–2780 (2010).
38. V. L. Bratman, A. W. Cross, G. G. Denisov, W. He, A. D. R. Phelps, K. Ronald, S. V. Samsonov, C. G. Whyte, and A. R. Young, "High-gain wide-band gyrotron traveling wave amplifier with a helically corrugated waveguide," Phys. Rev. Lett. **84**, 2746–2749 (2000).
39. E. A. Nanni, S. K. Jawla, M. A. Shapiro, P. P. Woskov, and R. J. Temkin, "Low-Loss Transmission Lines for High-Power Terahertz Radiation," J Infrared Millim Terahertz Waves **33**, 695–714 (2012).
40. K. Ohkubo, S. Kubo, H. Idei, M. Sato, T. Shimozuma, and Y. Takita, "Coupling of tilting Gaussian beam with hybrid mode in the corrugated waveguide," Int J Infrared Milli Waves **18**, 23–41 (1997).
41. J. Ye, L.-S. Ma, and J. L. Hall, "Ultrasensitive detections in atomic and molecular physics: demonstration in molecular overtone spectroscopy," J. Opt. Soc. Am. B, JOSAB **15**, 6–15 (1998).
42. T. P. Hua, Y. R. Sun, J. Wang, A. W. Liu, and S. M. Hu, "Frequency metrology of molecules in the near-infrared by NICE-OHMS," Opt. Express, OE **27**, 6106–6115 (2019).

43. J. Dai, J. Zhang, W. Zhang, and D. Grischkowsky, "Terahertz time-domain spectroscopy characterization of the far-infrared absorption and index of refraction of high-resistivity, float-zone silicon," J. Opt. Soc. Am. B, JOSAB **21**, 1379–1386 (2004).
44. J. L. Doane, "Design of Circular Corrugated Waveguides to Transmit Millimeter Waves at ITER," Fusion Science and Technology **53**, 159–173 (2008).
45. G. R. Hanson, J. B. Wilgen, T. S. Bigelow, S. J. Diem, and T. M. Biewer, "Analysis of the ITER low field side reflectometer transmission line system," Review of Scientific Instruments **81**, 10D920 (2010).
46. P. P. Woskov, V. S. Bajaj, M. K. Hornstein, R. J. Temkin, and R. G. Griffin, "Corrugated waveguide and directional coupler for CW 250-GHz gyrotron DNP experiments," IEEE Transactions on Microwave Theory and Techniques **53**, 1863–1869 (2005).
47. W. Withayachumnankul, B. M. Fischer, and D. Abbott, "Quarter-wavelength multilayer interference filter for terahertz waves," Optics Communications **281**, 2374–2379 (2008).
48. J. Lott, C. Xia, L. Kosnosky, C. Weder, and J. Shan, "Terahertz Photonic Crystals Based on Barium Titanate/Polymer Nanocomposites," Advanced Materials **20**, 3649–3653 (2008).
49. Y. Han, M. Cho, H. Park, K. Moon, E. Jung, and H. Han, "Terahertz Time-domain Spectroscopy of Ultra-high ReflectancePhotonic Crystal Mirrors," Journal of the Korean Physical Society **55**, 508–511 (2009).
50. P. Balzerowski, E. Bründermann, and M. Havenith, "Fabry–Pérot Cavities for the Terahertz Spectral Range Based on High-Reflectivity Multilayer Mirrors," IEEE Transactions on Terahertz Science and Technology **6**, 563–567 (2016).
51. C. P. Endres, S. Schlemmer, P. Schilke, J. Stutzki, and H. S. P. Müller, "The Cologne Database for Molecular Spectroscopy, CDMS, in the Virtual Atomic and Molecular Data Centre, VAMDC," Journal of Molecular Spectroscopy **327**, 95–104 (2016).
52. L. S. Rothman, I. E. Gordon, Y. Babikov, A. Barbe, D. Chris Benner, P. F. Bernath, M. Birk, L. Bizzocchi, V. Boudon, L. R. Brown, A. Campargue, K. Chance, E. A. Cohen, L. H. Coudert, V. M. Devi, B. J. Drouin, A. Fayt, J.-M. Flaud, R. R. Gamache, J. J. Harrison, J.-M. Hartmann, C. Hill, J. T. Hodges, D. Jacquemart, A. Jolly, J. Lamouroux, R. J. Le Roy, G. Li, D. A. Long, O. M. Lyulin, C. J. Mackie, S. T. Massie, S. Mikhailenko, H. S. P. Müller, O. V. Naumenko, A. V. Nikitin, J. Orphal, V. Perevalov, A. Perrin, E. R. Polovtseva, C. Richard, M. A. H. Smith, E. Starikova, K. Sung, S. Tashkun, J. Tennyson, G. C. Toon, Vl. G. Tyuterev, and G. Wagner, "The HITRAN2012 molecular spectroscopic database," Journal of Quantitative Spectroscopy and Radiative Transfer **130**, 4–50 (2013).
53. K. Justtanont, T. Khouri, M. Maercker, J. Alcolea, L. Decin, H. Olofsson, F. L. Schöier, V. Bujarrabal, A. P. Marston, D. Teyssier, J. Cernicharo, C. Dominik, A. de Koter, G. Melnick, K. M. Menten, D. Neufeld, P. Planesas, M. Schmidt, R. Szczerba, and R. Waters, "Herschel/HIFI observations of O-rich AGB stars: molecular inventory," A&A **537**, A144 (2012).
54. A. V. Burenin, A. N. Val'dov, E. N. Karyakin, A. F. Krupnov, and S. M. Shapin, "Submillimeter microwave spectrum and spectroscopic constants of the OCS molecule: Isotopic species $^{16}O^{12}C^{33}S$ and $^{18}O^{12}C^{32}S$," Journal of Molecular Spectroscopy **87**, 312–315 (1981).
55. A. Foltynowicz, F. M. Schmidt, W. Ma, and O. Axner, "Noise-immune cavity-enhanced optical heterodyne molecular spectroscopy: Current status and future potential," Appl. Phys. B **92**, 313 (2008).
56. N. Picqué and T. W. Hänsch, "Frequency comb spectroscopy," Nature Photonics **13**, 146 (2019).